\documentclass[preprint,showpacs,preprintnumbers,amsmath,amssymb,showkeys]{revtex4}
\usepackage{graphicx}% Include figure files
\usepackage{dcolumn}% Align table columns on decimal point
\usepackage{bm}% bold math

\begin{document}

\preprint{}

\title{Analytical vectorial structure of non-paraxial four-petal Gaussian
beams in the far field }% Force line breaks with \\
\author{Xuewen Long$^{a,b}$}
%\altaffiliation[ ]{}%Lines break automatically or can be forced with \\
\author{Keqing Lu$^{a}$}\email{keqinglu@opt.ac.cn}
\author{Yuhong Zhang$^{a,b}$}
\author{Jianbang Guo$^{a,b}$}
\author{Kehao Li$^{a,b}$}
%%\email{Second.Author@institution.edu}
\affiliation{$^{a}$State Key Laboratory of Transient Optics and
Photonics, Xi'an Institute of Optics and Precision Mechanics,
Chinese Academic of Sciences, Xi'an 710119, China
\\$^{b}$ Graduate school of Chinese Academy of Sciences, Beijing, 100039, China}
\date{\today}% It is always \today, today,

\begin{abstract}
The analytical vectorial structure of non-paraxial four-petal
Gaussian beams(FPGBs) in the far field has been studied based on
vector angular spectrum method and stationary phase method. In
terms of analytical electromagnetic representations of the TE and
TM terms, the energy flux distributions of the TE term, the TM
term,  and the whole beam are derived in the far field,
respectively. According to our investigation, the FPGBs can evolve
into a number of small petals in the far field. The number of the
petals is determined by the  order of input beam. The physical
pictures of the FPGBs are well illustrated from the vectorial
structure, which is beneficial to strengthen the understanding of
vectorial properties of the FPGBs.
\end{abstract}


%Use showkeys class option if keyword
\maketitle

\section{Introduction}\label{SecI}
Recently, a new form of laser beams called four-petal Gaussian
beams(FPGBs) has been introduced and its properties of passing
through a paraxial ABCD optical system have been
studied~\cite{Duan2006OC}. Subsequently, the propagation
properties of the FPGBs have attracted considerable interest due
to its potential applications. Gao and L\"{u} studied its
vectorial non-paraxial propagation in free space based on
vectorial Rayleigh-Sommerfeld diffraction integral formulas in
2006 ~\cite{Gao2006CPL}. In 2007, the propagation of four-petal
Gaussian beams in turbulent atmosphere was investigated by Chu
\emph{et al}~\cite{Chu2008CPL}. In 2008, Tang \emph{et al}
explored diffraction properties of four-petal Gaussian beams in
uniaxially anisotropic crystal by virtue of  the paraxially
vectorial theory of beam propagation~\cite{Tang2008COL}. Yang
\emph{et al} reported the propagation of four-petal Gaussian beams
in strongly nonlocal nonlinear media in the next
year~\cite{Yang2010OC}. In the meanwhile, the vectorial structure
of lots of beam with different patterns and polarized status  is
illustrated in the far field by means of vector angular spectrum
method~\cite{Rosario2001JOSAA}, which  is a useful tool to resolve
the Maxwell's equations, and stationary phase
method~\cite{Mandel,Born}, which uses the asymptotic approximation
approaching some kind of difficult integral. Based on these two
methods mentioned above, Zhou studied analytical vectorial
structure of Laguerre-Gaussian
 beam in the far field firstly~\cite{Zhou2006OL}. Afterwards, Deng and Guo explored
analytical vectorial structure of radially polarized light
beams~\cite{DengOL}. In 2008, Wu \emph{et al} and Zhou \emph{et
al} investigated vectorial structure of hollow Gaussian beam
almost in the same time~\cite{Wu2008OE,ZhouHollow2008OC}. In fact,
much work has been done  with  the vectorial structure of all
kinds of beams in the far
field~\cite{Zhounonsymmetrical2007OC}-~\cite{Zhou2009JOSAB}.

It is well known that the paraxial approximation is no longer
valid for beams with a large divergent angle or, especially, a
small beam spot size that is comparable with the light
wavelength~\cite{Nemoto}. Therefore, rigorous non-paraxial and
vectorial treatments are necessary. We can approach non-paraxial
propagation of beams in terms of  vector angular spectrum method
of electromagnetic field. According to vector angular spectrum
method of electromagnetic field, the general solution of the
Maxwell{$^{'}$s equations is  composed of the transverse electric
(TE) term and the transverse magnetic (TM) term. In the far field,
the TE and TM terms are orthogonal to each other and can be
detached.

To the best of our knowledge, the research on the vectorial
structure of non-paraxial four-petal Gaussian beam in the far
field based on vector angular spectrum method and stationary phase
method has not been reported elsewhere. In this paper, the
far-field vectorial properties  of non-paraxial four-petal
Gaussian beam have been studied by means of vector angular
spectrum method and stationary phase method. Based on the
analytical vectorial structure of the FPGBs, the energy flux
distributions of TE term, TM term and the whole FPGBs are also
investigated, respectively.

\section{Analytical vectorial structure in the far field}\label{SecII}
Let us consider a half space  $z>0$ filled with a linear
homogeneous, isotropic, nonconducting medium characterized by
electric permittivity $\varepsilon$ and magnetic permeability
$\mu$. All the sources only lie in the domain  $z<0$. The electric
field distribution is known at the boundary plane $z=0$. For
convenience of discussion, we consider a non-paraxial  FPGBs with
polarization in x direction, which propagates toward the half
space $z\geq0$ along the z axis. The initial transverse electric
field distribution of the FPGBs at the $z=0$ plane can be written
by~\cite{Duan2006OC}

\begin{equation}
E_x(x,y,0)=G_n\left(\frac{xy}{w_0^2}\right)^{2n}\exp\left(-\frac{x^2+y^2}{w_0^2}\right),
n=1,2,3\ldots\label{z0Ex},
\end{equation}
\begin{equation}
E_y(x,y,0)=0\label{z0Ey},
\end{equation}
where n is the beam order of the FPGBs; ${G_n}$ is a normalized
amplitude constant associated with the order of n; $w_0$ is the
 $1/e^2$ intensity waist radius of the Gaussian
term. The time factor $\exp(-i\omega t)$ has been omitted in the
field expression. Fig.~\ref{fig1} shows intensity distributions of
four-petal Gaussian beams at the initial plane $z = 0$ for $n = 1,
5, 9$ and 13, respectively. From Fig.~\ref{fig1}, it can be seen
that the intensity distributions is composed of four equal petals.
The distance of small petals increases when beam order increases.
Here, $w_0$ is taken by $\lambda$ in the calculation, which is
comparable with the wavelength. So, this is a non-paraxial
problem.

\begin{figure}[htbp]
\begin{center}
\includegraphics[width=12cm]{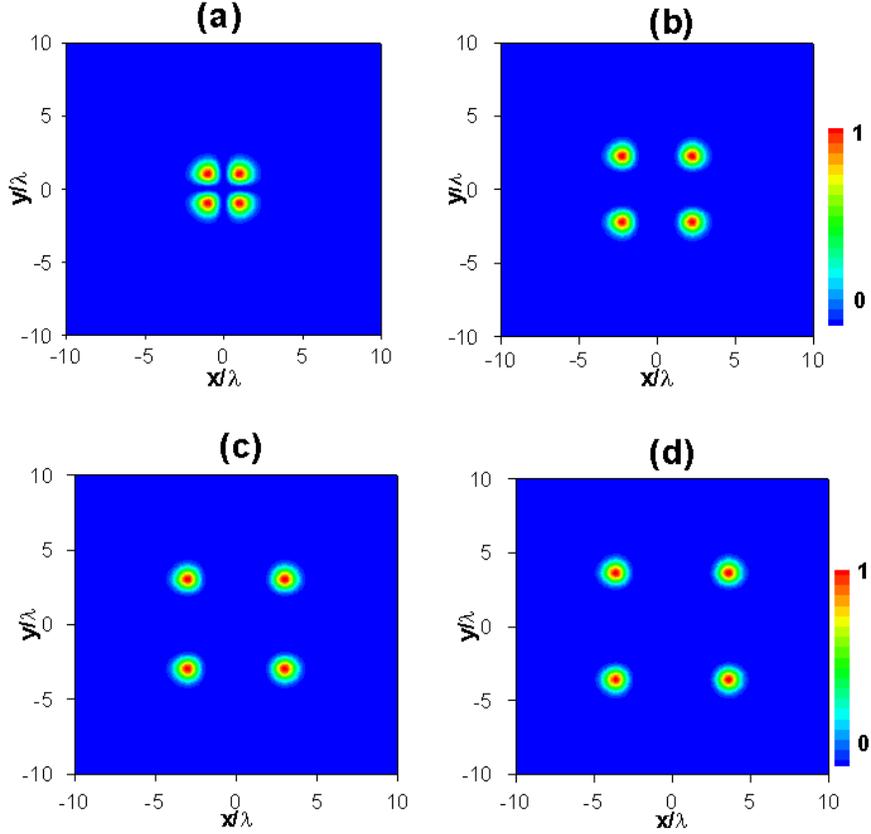}
\end{center}
\caption{\label{fig1}(Color online)Normalized intensity
distributions of FPGBs with different beam order n at $z=0$ plane
based Eq. (\ref{z0Ex}). (a) $n=1$, (b)
 $n=5$, (c) $n=9$, (d) $n=13$.}
\end{figure}

In terms of Fourier transform, the vectorial angular spectrum of
electric field at the $z=0$ plane is expressed as~\cite{Wu2008OE}
\begin{equation}
A_x(p,q)=\frac{1}{\lambda^2}\int\int^{\infty}_{-\infty}E_x(x,y,0)\exp[-ik(px+qy)]dxdy\label{Ax},
\end{equation}

\begin{equation}
A_y(p,q)=\frac{1}{\lambda^2}\int\int^{\infty}_{-\infty}E_y(x,y,0)\exp[-ik(px+qy)]dxdy\label{Ay},
\end{equation}
where $\lambda$ denotes the wave length in the medium related wave
number by $k=2\pi/\lambda$. Substituting Eqs. (\ref{z0Ex}) and
(\ref{z0Ey}) into Eqs. (\ref{Ax}) and (\ref{Ay}), we find that

\begin{eqnarray}
A_x(p,q)&=&\frac{G_n}{\lambda^2}w_0^2\left[\Gamma(n+\frac{1}{2})\right]^2
{}_1F_1\left(n+\frac{1}{2};\frac{1}{2};-\frac{1}{4}k^2p^2w_0^2\right)
 \nonumber \\
&&\times
_1F_1\left(n+\frac{1}{2};\frac{1}{2};-\frac{1}{4}k^2q^2w_0^2\right)\label{freespace},
\end{eqnarray}
where $_1F_1(\cdot;\cdot;\cdot)$ denotes confluent hypergeometric
function, $\Gamma(\cdot)$ denotes the Gamma function. It is well
known that Maxwell's equations can be separated into transverse
and longitudinal field equations and an arbitrary polarized
electromagnetic beam, which is expressed in terms of vector
angular spectrum, is composed of the transverse electric (TE) term
and the transverse magnetic (TM)term, namely,
\begin{eqnarray}
\vec{E}(\vec{r})=\vec{E}_{TE}(\vec{r})+\vec{E}_{TM}(\vec{r})\label{E},
\end{eqnarray}

\begin{eqnarray}
\vec{H}(\vec{r})=\vec{H}_{TE}(\vec{r})+\vec{H}_{TM}(\vec{r})\label{H},
\end{eqnarray}

where
\begin{eqnarray}
\vec{E}_{TE}(\vec{r})&=&\int\int^{\infty}_{-\infty}\frac{1}{p^2+q^2}[qA_x(p,q)-pA_y(p,q)](q\hat{e}_x-p\hat{e}_y)
 \nonumber \\
&&\times\exp(iku)dpdq\label{ETE},
\end{eqnarray}

\begin{eqnarray}
\vec{H}_{TE}(\vec{r})&=&\sqrt{\frac{\varepsilon}{\mu}}\int\int^{\infty}_{-\infty}\frac{1}{p^2+q^2}[qA_x(p,q)-pA_y(p,q)](p\gamma\hat{e}_x+q\gamma\hat{e}_y-b^2\hat{e}_z)
 \nonumber \\
&&\times\exp(iku)dpdq\label{HTE},
\end{eqnarray}
and
\begin{eqnarray}
\vec{E}_{TM}(\vec{r})&=&\int\int^{\infty}_{-\infty}\frac{1}{p^2+q^2}[pA_x(p,q)+qA_y(p,q)](p\hat{e}_x+q\hat{e}_y-\frac{b^2}{\gamma}\hat{e}_z)
 \nonumber \\
&&\times\exp(iku)dpdq\label{ETM},
\end{eqnarray}

\begin{eqnarray}
\vec{H}_{TM}(\vec{r})&=&-\sqrt{\frac{\varepsilon}{\mu}}\int\int^{\infty}_{-\infty}[pA_x(p,q)+qA_y(p,q)]\frac{1}{b^2\gamma}(q\hat{e}_x-p\hat{e}_y,)
 \nonumber \\
&&\times\exp(iku)dpdq\label{HTM}.
\end{eqnarray}
$\vec{r}=x\hat{e}_x+y\hat{e}_y+z\hat{e}_z$ is the displacement
vector and $\hat{e}_x$, $\hat{e}_y$, $\hat{e}_z$ denote unit
vectors in the x, y, z directions, respectively; $u=px+qy+\gamma
z$; $b^2=p^2+q^2$; $\gamma=\sqrt{1-p^2-q^2}$, if $p^2+q^2\leq1$ or
$\gamma=i\sqrt{p^2+q^2-1}$, if $p^2+q^2>1$. The value of
$p^2+q^2>1$ corresponds to the evanescent wave which propagates
along the boundary plane but decays exponentially along the
positive z direction.

In the far field framework, the condition
$k(x^2+y^2+z^2)^{1/2}\rightarrow\infty$ is satisfied due to z is
big enough. Moreover the contribution of the evanescent wave to
the far field can be ignored. By virtue of the method of
stationary phase~\cite{Mandel,Born,Wu2008OE}, the TE mode and the
TM mode of the electromagnetic field can be given by
\begin{eqnarray}
\vec{E}_{TE}(\vec{r})&=&-i\frac{G_nw_0^2}{\lambda}\frac{yz}{r^2\rho^2}\left[\Gamma(n+\frac{1}{2})\right]^2
{}_1F_1\left(n+\frac{1}{2};\frac{1}{2};-\frac{1}{4}k^2\frac{x^2}{r^2}w_0^2\right)
 \nonumber \\
&&\times{}_1F_1\left(n+\frac{1}{2};\frac{1}{2};-\frac{1}{4}k^2\frac{y^2}{r^2}w_0^2\right)\exp(ikr)(y\hat{e}_x-x\hat{e}_y)\label{Ete},
\end{eqnarray}

\begin{eqnarray}
\vec{H}_{TE}(\vec{r})&=&-i\sqrt{\frac{\varepsilon}{\mu}}\frac{G_nw_0^2}{\lambda}\frac{yz}{r^3\rho^2}\left[\Gamma(n+\frac{1}{2})\right]^2
{}_1F_1\left(n+\frac{1}{2};\frac{1}{2};-\frac{1}{4}k^2\frac{x^2}{r^2}w_0^2\right)
 \nonumber \\
&&\times_1F_1\left(n+\frac{1}{2};\frac{1}{2};-\frac{1}{4}k^2\frac{y^2}{r^2}w_0^2\right)\exp(ikr)
 \nonumber \\
&&\times(xz\hat{e}_x+yz\hat{e}_y-\rho^2\hat{e}_z)\label{Hte},
\end{eqnarray}

and
\begin{eqnarray}
\vec{E}_{TM}(\vec{r})&=&-i\frac{G_nw_0^2}{\lambda}\frac{x}{r^2\rho^2}\left[\Gamma(n+\frac{1}{2})\right]^2
{}_1F_1\left(n+\frac{1}{2};\frac{1}{2};-\frac{1}{4}k^2\frac{x^2}{r^2}w_0^2\right)
 \nonumber \\
&&\times_1F_1\left(n+\frac{1}{2};\frac{1}{2};-\frac{1}{4}k^2\frac{y^2}{r^2}w_0^2\right)\exp(ikr)
 \nonumber \\
&&\times(xz\hat{e}_x+yz\hat{e}_y-\rho^2\hat{e}_z)\label{Etm},
\end{eqnarray}

\begin{eqnarray}
\vec{H}_{TM}(\vec{r})&=&i\sqrt{\frac{\varepsilon}{\mu}}\frac{G_nw_0^2}{\lambda}\frac{x}{r\rho^2}\left[\Gamma(n+\frac{1}{2})\right]^2
{}_1F_1\left(n+\frac{1}{2};\frac{1}{2};-\frac{1}{4}k^2\frac{x^2}{r^2}w_0^2\right)
 \nonumber \\
&&\times_1F_1\left(n+\frac{1}{2};\frac{1}{2};-\frac{1}{4}k^2\frac{y^2}{r^2}w_0^2\right)\exp(ikr)
 \nonumber \\
&&\times(y\hat{e}_x-x\hat{e}_y)\label{Htm},
\end{eqnarray}
where $\rho=\sqrt{x^2+y^2}$ ; $r=\sqrt{x^2+y^2+z^2}$. Eqs.
(\ref{Ete})- (\ref{Htm}) are analytical vectorial expressions for
the TE and TM terms in the far field and constitute the basic
results  in this paper. From Eqs. (\ref{Ete})- (\ref{Htm}), we
find that

\begin{eqnarray}
\vec{E}_{TE}(\vec{r})\cdot\vec{E}_{TM}(\vec{r})=0\label{ETEETM},
\end{eqnarray}
\begin{eqnarray}
\vec{H}_{TE}(\vec{r})\cdot\vec{H}_{TM}(\vec{r})=0\label{HTEHTM}.
\end{eqnarray}
 According to Eqs. (\ref{ETEETM}) and (\ref{HTEHTM}), the TE and TM terms
of FPGBs are orthogonal to each other in the far field.

\section{Energy flux distributions in the far field}\label{SecIII}
The energy flux distributions of the TE and TM terms at the
$z=const $ plane are expressed in terms of the z component of
their time-average Poynting vector as
\begin{eqnarray}
\langle
S_z\rangle_{TE}=\frac{1}{2}Re[\vec{E}_{TE}^*\times\vec{H}_{TE}]_z{}\label{SZTE},
\end{eqnarray}
\begin{eqnarray}
\langle
S_z\rangle_{TM}=\frac{1}{2}Re[\vec{E}_{TM}^*\times\vec{H}_{TM}]_z{}\label{SZTM},
\end{eqnarray}
where the Re denotes real part, and the asterisk denotes complex
conjugation. The whole energy flux distribution of the beam is a
sum of the energy flux of TE mode and TM mode, namely,
\begin{eqnarray}
\langle S_z\rangle=\langle S_z\rangle_{TE}+\langle
S_z\rangle_{TM}\label{SZ},
\end{eqnarray}
Substituting Eqs. (\ref{Ete})- (\ref{Htm}) into Eqs. (\ref{SZTE})-
(\ref{SZTM}) yields

\begin{eqnarray}
\langle
S_z\rangle_{TE}&=&\frac{1}{2}\sqrt{\frac{\varepsilon}{\mu}}\frac{G_n^2w_0^4}{\lambda^2}\frac{y^2z^3}{r^5\rho^2}\left[\Gamma(n+\frac{1}{2})\right]^4
{}_1F_1\left(n+\frac{1}{2};\frac{1}{2};-\frac{1}{4}k^2\frac{x^2}{r^2}w_0^2\right)^2
 \nonumber \\
&&\times_1F_1\left(n+\frac{1}{2};\frac{1}{2};-\frac{1}{4}k^2\frac{y^2}{r^2}w_0^2\right)^2\label{SzTEexpress},
\end{eqnarray}

\begin{eqnarray}
\langle
S_z\rangle_{TM}&=&\frac{1}{2}\sqrt{\frac{\varepsilon}{\mu}}\frac{G_n^2w_0^4}{\lambda^2}\frac{x^2z}{r^3\rho^2}\left[\Gamma(n+\frac{1}{2})\right]^4
{}_1F_1\left(n+\frac{1}{2};\frac{1}{2};-\frac{1}{4}k^2\frac{x^2}{r^2}w_0^2\right)^2
 \nonumber \\
&&\times_1F_1\left(n+\frac{1}{2};\frac{1}{2};-\frac{1}{4}k^2\frac{y^2}{r^2}w_0^2\right)^2\label{SzTMexpress},
\end{eqnarray}
Therefore, the whole energy flux distribution of FPGBs in the far
field is given by
\begin{eqnarray}
\langle
S_z\rangle&=&\frac{1}{2}\sqrt{\frac{\varepsilon}{\mu}}\frac{G_n^2w_0^4}{\lambda^2}\frac{z}{r^3\rho^2}\left(\frac{y^2z^2}{r^2}+x^2\right)\left[\Gamma(n+\frac{1}{2})\right]^4
\nonumber \\
&&\times{}_1F_1\left(n+\frac{1}{2};\frac{1}{2};-\frac{1}{4}k^2\frac{x^2}{r^2}w_0^2\right)^2
 \nonumber \\
&&\times{}_1F_1\left(n+\frac{1}{2};\frac{1}{2};-\frac{1}{4}k^2\frac{y^2}{r^2}w_0^2\right)^2\label{wholeenergyflux}.
\end{eqnarray}

\begin{figure}[htbp]
\begin{center}
\includegraphics[width=13cm]{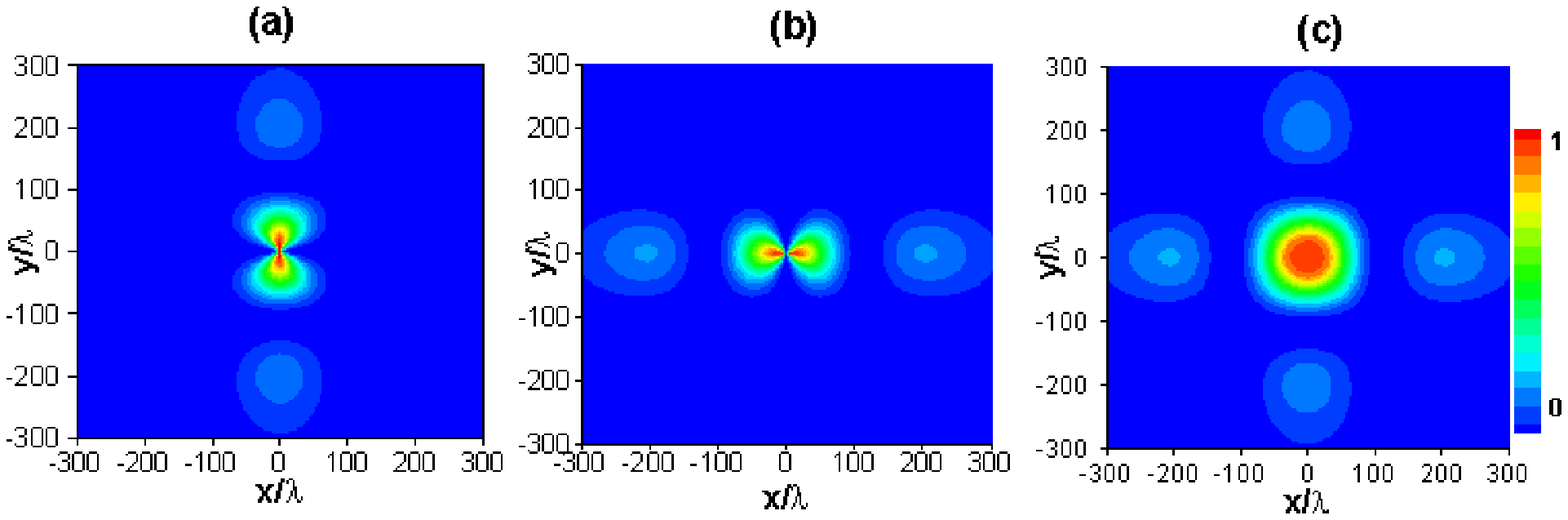}
\end{center}
\caption{\label{fig2}(Color online)Normalized energy flux
distributions of FPGBs at the plane $z=500\lambda$ for beam order
$n=1$. (a) The TE term, (b) The TM term, (c) The whole beam. }
\end{figure}

\begin{figure}[htbp]
\begin{center}
\includegraphics[width=13cm]{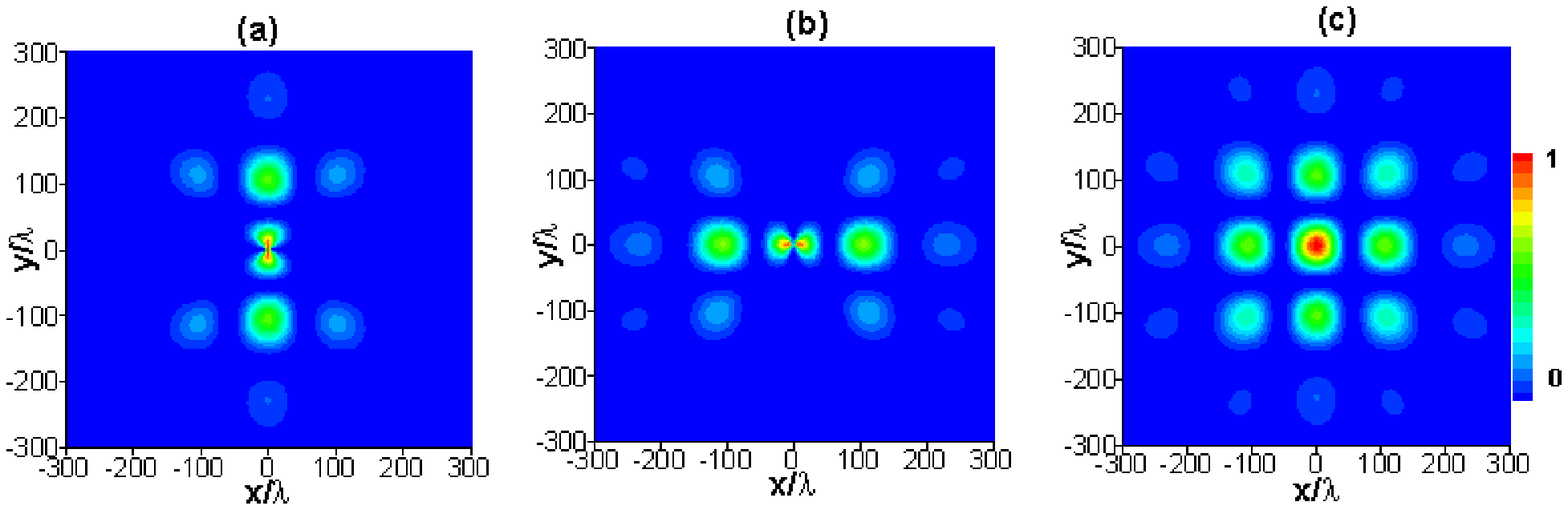}
\end{center}
\caption{\label{fig3}(Color online)Normalized energy flux
distributions of FPGBs at the plane $z=500\lambda$ for beam order
$n=5$. (a) The TE term, (b) The TM term, (c) The whole beam.}
\end{figure}

\begin{figure}[htbp]
\begin{center}
\includegraphics[width=13cm]{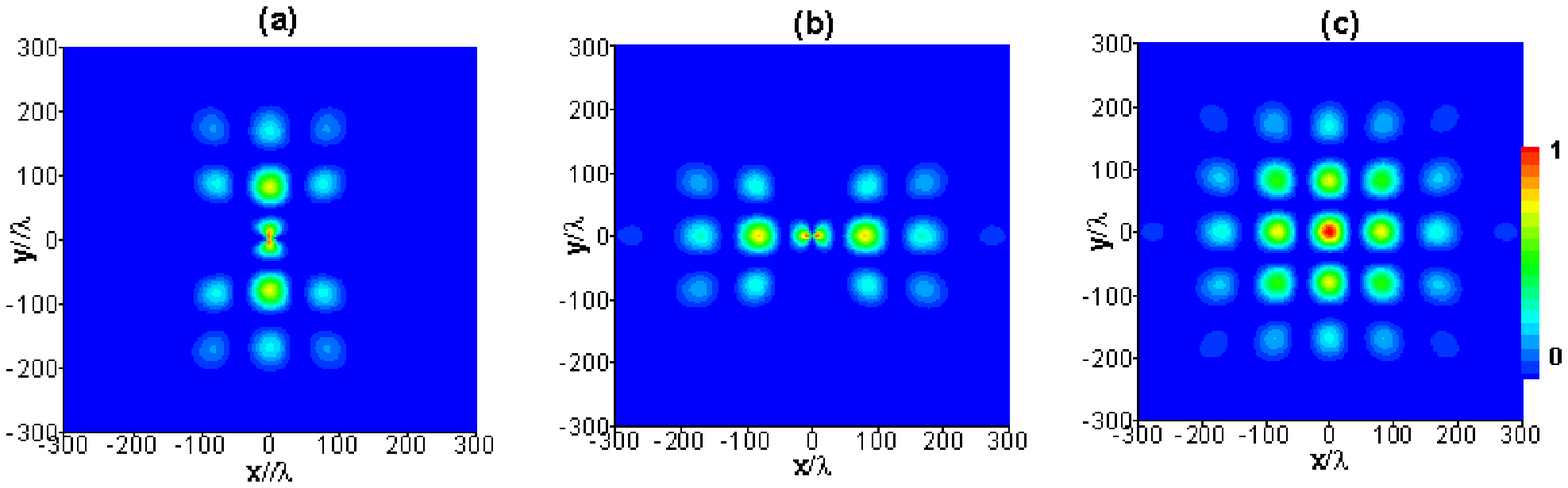}
\end{center}
\caption{\label{fig4}(Color online)Normalized energy flux
distributions of FPGBs at the plane $z=500\lambda$ for beam order
$n=9$. (a) The TE term, (b) The TM term, (c) The whole beam. }
\end{figure}

\begin{figure}[htbp]
\begin{center}
\includegraphics[width=13cm]{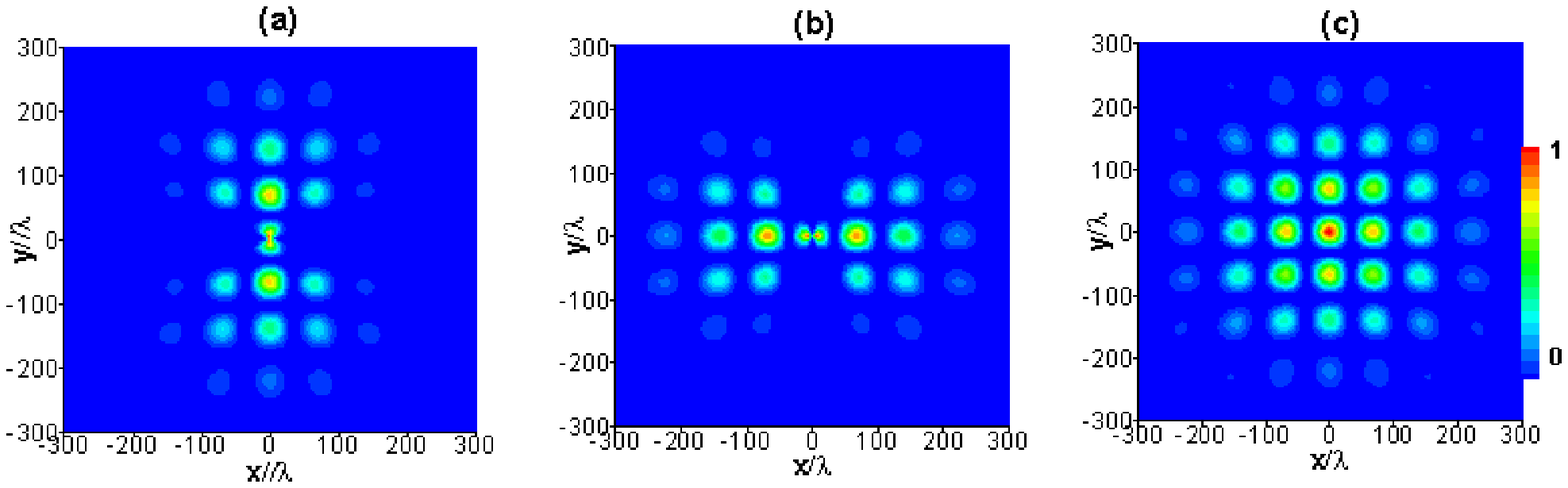}
\end{center}
\caption{\label{fig5}(Color online)Normalized energy flux
distributions of FPGBs at the plane $z=500\lambda$ for beam order
$n=13$.  (a) The TE term,(b) The TM term, (c) The whole beam.}
\end{figure}
The normalized energy flux distributions of the TE term, the TM
term and  the FPGBs at the plane  $z=500\lambda$  for different
beam order $n=1, 5, 9, 13$ versus $x/\lambda$ and $y/\lambda$ are
illustrated by Figs.~\ref{fig2}-~\ref{fig5}. The used parameter is
$w_0=\lambda$. As can be seen from  Figs. ~\ref{fig2}-~\ref{fig5},
the FPGBs split into a number of small petals in the far field,
which differs from its initial four-petal shape. The FPGBs with
beam order n is not a pure mode, which  can be regarded as a
superposition of $n^2$ two dimensional Hermite-Gaussian
modes~\cite{Duan2006OC}, and different modes evolve differently
within the same propagation distance. The overlap and interference
in propagation between different modes result in the propagation
properties of the FPGBs in the far field. Furthermore, the number
of petals is determined by the parameter n. The number of petals
in the far field gradually increases when the parameter n
increases, which has potential applications in micro-optics  and
beam splitting techniques, etc. Note that diameter of the central
beam spot decreases when beam order n increases. This phenomenon
has been discussed in some previous
researches~\cite{Duan2006OC}-~\cite{Chu2008CPL}.

\section{Conclusions}
In summary, the vectorial structure  of the non-paraxial
four-petal Gaussian beam in the far field is expressed in the
analytical form by using the vector angular spectrum method and
the  stationary phase method. The electric field and the magnetic
field of the four-petal Gaussian beam is decomposed into two
mutually orthogonal terms, i.e., TE term and TM term. Based on the
analytical vectorial structure of FPGBs, the energy flux
distributions of the TE term, the TM term and the whole beam of
FPGBs are derived in the far-field  and are illustrated by
numerical examples. The number of petals and diameter of the
central beam spot in the far field are determined by the beam
order n. The potential applications of the FPGBs are deserved
investigation. This work is important to understand the
theoretical aspects of vector FPGBs propagation.

\begin{acknowledgements}
This research was supported by the National Natural Science
Foundation of China (Grant No.10674176 ). The author is beneficial
from the discussion with the author of Ref.~\cite{Wu2008OE}.
\end{acknowledgements}

\end{document}